\documentclass[12pt]{article}

\usepackage{amssymb}
\usepackage{mathdots}
\usepackage{amsthm}
\usepackage{amscd}
\usepackage{amsfonts}
\usepackage{dsfont}
\usepackage{cite}
\usepackage{graphicx}
\usepackage{color}
\usepackage{multicol}
\usepackage{colortbl}
\usepackage{epsfig}
\usepackage{indentfirst}
\usepackage{longtable}
\usepackage{pstricks}
\usepackage{pst-all}
\usepackage[latin1]{inputenc}
\usepackage[english]{babel}

\def\be{\begin{equation}   }
\def\ee{\end{equation}   }
\def\ba{\begin{eqnarray}   }
\def\ea{\end{eqnarray}   }

\def\vp{\varphi}

\def\ra{\rightarrow}

\def\ni{\noindent}
\def\nn{\nonumber}

\def\ni{\noindent}
\def\pt{$\mathcal{PT}$}

\def\){\right)}
\def\({\left(}

\def\ele{2\ell+1}
\def\elef{\frac{2\ell+1}{2}}

\newcommand{\fl}{\hspace*{-0.5cm}}

\usepackage{graphicx}

\begin{document}

\begin{titlepage}
\vspace*{-1cm}

\vskip 3cm

\vspace{.2in}
\begin{center}
{\large\bf $A_{(2|1)}$ spectral equivalences and nonlocal integrals of motion}
\end{center}

\vspace{.5cm}

\begin{center}
P. E. G.  Assis

\vspace{.5 in}
\small

\par \vskip .2in \noindent
Scuola Internazionale Superiore di Studi Avanzati \\
and \\
Istituto Nazionale di Fisica Nucleare, sezione di Trieste, \\
Via Bonomea 265, Trieste, Italy \\
\vspace{1cm}
paulo.assis@sissa.it\\

\normalsize
\end{center}

\vspace{.5in}

\begin{abstract}

We study the spectral correspondence between a particular class of Schr\"{o}dinger equations and supersymmetric quantum integrable model (QIM). The latter, a quantized version of the  Ablowitz-Kaupp-Newell-Segur (AKNS) hierarchy of nonlinear equations, corresponds to the thermodynamic limit of the Perk-Schultz lattice model. By analyzing the symmetries of the ordinary differential equation (ODE) in the complex plane, it is possible to obtain important objects in the quantum integrable model in exact form, under an exact spectral correspondence. In this manuscript our main interest lies on the set of nonlocal conserved integrals of motion associated to the integrable system and we provide a systematic method to compute their values evaluated on the vacuum state of the quantum field theory.

\vspace{1cm}

\small
\textbf{Pacs:} 02.30.Hq, 02.30.Ik, 02.30.Jk, 02.30.Mw, 02.30.Rz, 02.60.Lj, 02.70.Hm, \\
\vspace{-0.5cm}
\qquad\qquad\, 02.70.Rr, 03.65-w, 03.65.Db, 03.65.Ge, 03.70.+k, 05.30.-d, 75.10.Pq.
\normalsize

\end{abstract}

\end{titlepage}

\setcounter{equation}{0}

\section{Introduction}

The existence of common symmetries underlying both ordinary differential equations and quantum integrable models was first observed for the simple anharmonic oscillator Schr\"{o}dinger potential in \cite{DT-Anharmonic}, from which functional relations characteristic to quantum integrable models were constructed. Soon after, in \cite{BLZ-SpecDet}, the addition of a centrifugal  term to the differential problem was shown to correspond to a twist in the boundary conditions of the lattice model. Moreover, this allowed to relate a differential eigenvalue problem to more general values of the Fock vacuum parameter in the QIM. The extension to excited states of the Virasoro module was achieved in \cite{BLZ-Higher} by introducing a singular part to the potential with poles located away from the origin. The overall symmetry of the differential problem in all cases, although hidden, is encoded by the $A_1 = su(2)$ Lie algebra. The quantum field theory associated to this symmetry has been thoroughly studied in e.g. \cite{BLZ-1, BLZ-2, BLZ-3}. Besides the twisted XXZ spin chain, also the 6-vertex lattice model, the massless twisted sine-Gordon integrable field theory and can be shown to be spectrally equivalent to this Schr\"{o}dinger equation.

The crucial ingredients in both frameworks are the Symanzik symmetry of rotations on the complex plane in the case of the ODE and the existence of a Yang-Baxter structure for the QIM. These two elements, although not directly related, lead to equivalent functional relations - the Bethe Ansatz Equation, the $T$-$Q$ or the Quantum Wronskian relations  - which allow one to establish a parallel between the transfer matrix $T(E)$ and Baxter's $Q(E)$ operator to spectral determinants $D(E)$ and Wronskians $W(E)$ of solutions to the O.D.E., e.g., \cite{DDT-Review}. Consequently, the eigenvalues of $Q(E)$ have zeros at the same points of the spectral parameter $E$ as $D(E)$. Thus, lattice models, interesting from the statistical viewpoint, and Integrable Quantum Field Theories (IQFT), possibly minimal models in Conformal Field Theories (CFT), can be conveniently mapped onto ordinary differential equations defined on the complex plane.

The problem in discussion is not only closely related to physical applications
but also has connections to more purely mathematical questions such as number theory. In fact, the eigenvalues of this problem can be combined in spectral Zeta functions, e.g. \cite{Voros}, allowing one to obtain some of the nonlocal charges \cite{Watkins}. With this in mind, the ODEs and integrable models can therefore be used to infer information about the Zeta functions.
Another important byproduct of the investigation on the aforementioned correspondence came in \cite{DDT-Reality}, with what is generally considered a reality proof for the spectra of a class of non-Hermitian \pt-symmetric Hamiltonians \cite{BenderBoettcher,BenderBoettcherMeisinger}. Therefore, the study of spectral equivalences between ODEs and integrable models found fertile ground on the \pt-symmetry community. 
This area received a great deal of attention from a theoretical point of view 
and more applications of the so called IM/ODE correspondence to $\mathcal{PT}$-symmetric Quantum Mechanics are natural.

In order to investigate the connection of the integrals of motion in the QIM to the differential problem, a simpler extension consists of utilizing the supersymmetric $A_{(2|1)} = su(2|1)$ Lie algebra. The advantage of this relies on the existence of a well understood quantum field theory, namely the AKNS soliton hierarchy \cite{FL-AKNS, BT-Susy}. In this situation we have close links to the Integrable Perturbed Hairpin model and the
Perk-Schultz model. Another positive aspect is that the $su(2|1)$ algebra, being one of the simplest supersymmetric extension of the Lie algebra, provides a good environment to test ideas in supersymmetry and superconformal field theory, which could play an important role in the fundamental descriptions of nature, according to AdS/CFT correspondence \cite{AdSCFT}.

In what follows, in order to fix the notation, we will present the class of ODEs of interest and review how to construct, from its symmetries, spectral functional relations. 
With a few identification we can show that this expansion corresponds to that associated to the quantum AKNS field theory and that the coefficients in this series are the eigenvalues of its local integrals of motion $I_n$. The nonlocal integrals of motion $\tilde{H}_n$, usually more difficultly obtained, are constructed through a novel perturbative analysis of the ODE after a suitable change of variables and the first few are explicitly presented.

\section{The Schr\"{o}dinger problem and functional relations}

Interested in studying an ordinary differential equation which could encode all the information about the quantum AKNS field theory, we turn our attentions for the moment, to solving the following differential equation,
\begin{eqnarray}\label{intperthairpinode}
\left( - \frac{d^2}{d y^2} +4 \left(P^2 + \bar{P} \kappa e^y \right)+ \kappa^2 \left( e^{2y} + e^{2 \beta^2 y} \right) \right) \Phi(y) =0,
\end{eqnarray}

\ni defined originally on the whole real line $-\infty<x<\infty$ for a restricted interval of the coupling $0 \leq \beta^2 \leq \frac{1}{2}$.
This is the equation presented in \cite{FL-AKNS} if we identify the parameters there as $p = 2 \imath P , q = 2\bar{P}, n = -2\beta^2$.
For our convenience we introduce a preliminary change of variables
\begin{eqnarray}
y = \log\left((-E)^{-\frac{M+1}{2M}}x^{M+1}\right), \qquad \qquad \Phi(y) = \frac{(M+1)^2}{\sqrt{y}} \psi(y),
\end{eqnarray}

\ni and a consort reparameterization given by
\begin{eqnarray}\label{aknsparamap}
\fl
P =  \frac{\ele}{4(M+1)}, \quad\quad
\bar{P} = \frac{\alpha}{4(M+1)}, \quad\quad
\beta^2=\frac{1}{M+1}, \quad\quad
\kappa = \frac{(-E)^{\frac{M+1}{2M}}}{M+1}.
\end{eqnarray}

This allows one to re-express the so-called Integrable Perturbed Hairpin model ODE (\ref{intperthairpinode}) in a more familiar fashion, namely, a generalized harmonic oscillator,
\begin{eqnarray}\label{genharmoscode}
\left( - \frac{d^2}{d x^2} + { \frac{\ell(\ell+1)}{x^2} } + {x^{2M}} + {\alpha x^{M-1}} - E \right) \psi(x) =0,
\end{eqnarray}

\ni defined on the half-line only, $0<x<\infty$, within the range $M \geq 1$, for which the harmonic oscillator $M=1$ is a limiting case.
The appearance of the $\alpha$-term was shown in \cite{Suzuki-FuncRel} to correspond to breaking the $su(2)$ symmetry into $su(2|1)$, for reasons which will become  clearer later.
The complete knowledge about the spectrum $\{ E_k\}$ can give information about the ground state of the associated quantum integral model.
It is also possible to extend (\ref{genharmoscode}) once more so that the excited states can also be accessed by the introduction of singularities at $x^{2(M+1)}=z_k$ in the potential \cite{BLZ-Higher, AD-Exc}.

The equation above is an extension of the $x^{2M}$ potential, or of the \pt-symmetric non-Hermitian
$(\imath x)^{2M}$ problem \cite{BenderBoettcher,BenderBoettcherMeisinger}, so that when $x \ra i x$ the associated eigenfunctions and eigenvalues are given by
$\tilde{\psi}(x) = \psi(i x), \lambda = -E$. 
The investigation of this family of potentials has generated a number of works on non-Hermitian Quantum Mechanics and for the non-supersymmetric problem, $\alpha=0$, 
the reality proof for its spectra was given in \cite{DDT-Reality}, whereas metrics and Hermitian counterparts for the $-x^4$-potential is found in \cite{JonesMateo}. More general non-Hermitian Hamiltonians have been studying outside the scope of the IM/ODE correspondence in e.g. \cite{FringFaria-Moyal, AF-Lie, A-SU2}.
It is also interesting to note that when taking $x \ra \imath x$, the part of the potential which has singularities away from the origin changes by moving the position of the poles,
$z_k \ra e^{-2 \pi \imath (M+1)} z_k$. In the event of $M$ being a half-integer the position of the poles change sign, whereas for an integer value they remain unchanged, being still closely related to those of the original problem.

The limiting $M=1$ case usually requires special attention as some of the results no longer hold and have to be modified. However, the introduction of an $\alpha \neq 0$ term has a unifying character, as the energy can be interpreted to be rescaled by an ammount $\alpha$, in this case. Therefore, the usual re-adaptation one has to do when evaluating the extended harmonic oscillator can be easily obtained by simply taking $\alpha = -E$,
\begin{eqnarray}\label{solE0a0M1}
\psi (x; M = 1,\ell ,\alpha -E , E= 0) &=& \psi (x; M =1,\ell ,\alpha ,E), \\
\nonumber
\psi (x; M = 1,\ell ,\alpha=-E , E= 0) &=& \psi (x; M =1,\ell ,\alpha=0 ,E).
\end{eqnarray}
This model ($M=1, \alpha=0$) is sometimes denoted the Hooke atom, because it describes a particle in the presence of  a centrifugal form alongside a linear Hooke force, is a benchmark for being exactly solvable. In the integrable models framework, this corresponds to a coupling of $\beta^2=\frac{1}{2}$, known as the free fermion point because in this situation the corresponding sine-Gordon field theory can be mapped onto a non-interacting massive Thirring model.
The solution of the Schr\"{o}dinger problem is given in terms of Laguerre polynomials and Hypergeometric functions, depending on the boundary conditions and eigenvalues, weighted by a decaying exponential factor. For general $M \neq 1$ the quantization problem cannot be solved exactly in closed form unless $E=0$. In this case,
\begin{eqnarray}\label{exactsolutionE0}\label{solE0}
\fl\qquad
\nonumber
\psi (M,\ell, \alpha ,E = 0,x)
&\propto& x^{\frac{1}{2} (1\pm (2 \ell +1))} e^{-\frac{x^{M+1}}{M+1}}
L_{-\frac{M+1\pm (2 \ell +1)+\alpha }{2 (M+1)}}^{\frac{2 \ell+1}{M+1}}\left(\frac{2 x^{M+1}}{M+1}\right),\\
\fl\qquad
\psi (M,\ell, \alpha ,E = 0,x) \label{exactsolE0}
&\propto& x^{\frac{1}{2} (1\pm (2 \ell +1))} e^{-\frac{x^{M+1}}{M+1}}
U_{\frac{M+1\pm (2 \ell+1)+\alpha}{2 (M+1)}}^{\frac{M+1\pm (2 \ell +1)}{M+1}}\left(\frac{2 x^{M+1}}{M+1}\right),
\end{eqnarray}
where $L$ and $U$ denote Laguerre and Hypergeometric functions.

Still in the harmonic oscillator regime, we can make use of the fact that the differential equation is also invariant under
$x \ra i x, \quad \alpha \ra -\alpha, \quad E \ra -E$, which can be seen as a $\mathcal{PT}$-symmetrization of the problem. In fact, more generally, equation (\ref{genharmoscode}) is invariant under discrete rotations by multiples of a certain angle on the complex plane, alongside a change in the constants $\alpha$ and $E$. The so-called Symanzik transformations for this system read \cite{Sibuya}
\begin{eqnarray}\label{ptsymzation}
\fl \qquad
x \; \rightarrow \; \omega^{-k} x, \quad
\alpha \rightarrow  (-1)^{k}\alpha, \quad
E \; \rightarrow \;  \omega^{2k}E, \quad
~\omega = \exp \( \frac{\imath \pi}{M+1} \).
\end{eqnarray}
Since symmetries in general help one solving differential equations, we can use the this invariance to generate new solutions, suitably normalised,
\begin{eqnarray}
\fl\qquad
\psi_k (x; M,\ell,E,\alpha) \equiv (-1)^{k+1}\imath \, \omega^{\frac{k}{2}(1-e^{\imath k \pi}\alpha)} \psi (\omega^{-k}x;M,\ell,\omega^{2k}E, e^{\imath k \pi}\alpha),
\end{eqnarray}

\ni which solve the ODE for any $k \in \mathds{Z}$.
It is the unique subdominant solution, specified by the condition that it decays exponentially at infinity, in the region $S_k$,
\begin{eqnarray}\label{sectorSk}
S_k = \left\{ \left| \textrm{arg}(x) - \frac{k\pi}{M+1} \right| < \frac{\pi}{2(M+1)} \right\}.
\end{eqnarray}

Once one knows a solution of the second order differential equation, the symmetry on the complex plane can be used to generate a new linearly independent solution, and one can use these two to write any solution of the problem.
However, allowing (\ref{genharmoscode}) in the complex plane requires a consistent choice of boundary conditions, as square-integrable wavefuntions can only exist in appropriate regions. 
The Stokes sectors in the complex plane define wedges where eigenfunctions exponentially decay and blow up in an alternating pattern. 
The problem we are mainly interested in throughout this manuscript is the radial problem, defined on the positive real axis and imposed to vanish at the origin and infinity, within the sector $S_0$. 


The $k$-th rotated solution, $\psi_k$ is subdominant in $S_k$ and dominant in $S_{k+1}$. Moreover, $\{\psi_k, \psi_{k+1} \} $ forms a basis of solution and $W[\psi_{k-1}, \psi_{k+1}]$ will vanish if and only if there exists a solution decaying in both sectors.
Thus, the solution $\psi_0=\psi$ one may find for the problem is, for instance, linearly independent to $\psi_{\pm 1}$, solutions on the complex plane rotated by an angle $\omega^{\pm 1}$ with respect to $\psi_0$. In our problem, defined on the positive real line, with vanishing boundary conditions at the origin and at infinity, $\psi_0$ is defined on the real axis, whereas $\psi_{\pm 1}$ satisfy boundary conditions on certain regions of the complex plane. To be more precise, the appropriate boundary condition at infinity for our radial problem implies it decays exponentially according to
\begin{eqnarray}\label{asymPsi}
\psi(x; M,\ell,E,\alpha) \sim \Psi (x; M,\ell,E,\alpha) \equiv x^{-\frac{M+\alpha}{2}}\exp\left( -\frac{x^{M+1}}{M+1} \right).
\end{eqnarray}
Note that when $M$ is not an integer $\mathds{N}$, the potential $V(x)$ is not single valued, so a branch cut, chosen along the negative real axis, is necessary.




Another symmetry of the general ordinary differential equation (\ref{genharmoscode}) is that it is invariant under $\ell \rightarrow -(\ell+1)$. Therefore, there are two linearly independent solutions,
\begin{eqnarray}
\psi^+ (x; M,E,\alpha) & \equiv & \psi (x; M,\ell,E,\alpha), \\
\nonumber
\psi^- (x; M,E,\alpha) & \equiv & \psi (x; M,-(\ell+1),E,\alpha).
\end{eqnarray}
The parameter $\ell$  dictates the behaviour near the origin and the expressions above indicate that, although (\ref{exactsolE0}) has a singular behaviour near the origin of the type $x^{-\ell}$, it can be easily analytically continued to a regular solution. Thus, by simply using this transformation one recovers square integrability, an important property when considering the Hermiticity of the differential problem.
Near the origin the solutions behave like
\begin{eqnarray}\label{basisorigin}
\psi^{\pm}(x; M,\ell,E,\alpha) \sim \vp^{\pm} (x;M,\ell,E,\alpha) \equiv x^{\frac{1}{2} \pm \frac{\ele}{2}}, \quad \textrm{as} \quad x \rightarrow 0,
\end{eqnarray}
and one can use them as a basis of expansion for a solution in this region.
Because the problem is not solved exactly in general for $M \neq 1$ one must instead resort to constructing solutions near the origin by the Frobenius method and using the WKB method to find asymptotic solutions which can furnish us with non-perturbative information about its analytical structure.

The basis of solutions $\vp^\pm$ around the origin (\ref{basisorigin}) form a linearly independent set satisfying the following Wronskian identities,
\begin{eqnarray}
W_{[-,+]} = W[\vp^-(x; M,\ell,E,\alpha),\vp^+(x; M,\ell,E,\alpha)] =  \ele,
\end{eqnarray}
\begin{eqnarray}\label{quantwronele}
\nonumber
(\ele)\psi (x; M,\ell,E,\alpha) & = &
W_{[0,+]}(M,\ell,E,\alpha) {\vp^- (x; M,\ell,E,\alpha)} +\\
&+& W_{[0,-]}(M,\ell,E,\alpha) {\vp^+ (x; M,\ell,E,\alpha)},
\end{eqnarray}
with $x$-independent coefficients given by the Wronskians $W_{[0,\pm]} =  W[\psi, \varphi^\pm]$ of a solution $\psi_0 (x; M,\ell,E,\alpha) $ with this pair, so that  one can use them to expand a general solution in this region. 
The exact relation between them is given in terms of Wronskians and because $W_{[0,1]} = W_{[-1,0]}$ we can use a normalization condition of $\psi_k$ so that
\begin{equation}
\fl
W_{[-1,1]}(M,\ell,E,\alpha) \psi_{0} (x; M,\ell,E,\alpha) =
\psi_{-1} (x; M,\ell,E,\alpha)+  \psi_{1} (x; M,\ell,E,\alpha),
\end{equation}
and in order to eliminate completely the $x$-dependence of this relation, one takes the Wronskian of this equation with the pair $\vp^\pm$, giving
\begin{eqnarray}
\fl
\nonumber
W_{[-1,1]}(M,\ell,E,\alpha) W_{[0,\pm]}(M,\ell,E,\alpha) &=&
\omega^{-\frac{\alpha \pm (\ele)}{2}}W_{[0,\pm]} (M, \ell, \omega^{-2}E,-\alpha) +  \\
\fl
&+& \omega^{+\frac{\alpha \pm (\ele)}{2}}W_{[0,\pm]} (M, \ell, \omega^{+2}E,-\alpha).
\end{eqnarray}

One of the quantities in the left hand side involves the asymptotic behaviour in the sectors $S_{\pm 1}$ of the complex plane and is therefore associated to the lateral problem and the \pt-symmetric differential equation; namely, $W_{[-1,1]}$ vanishes if and only if, for a certain value $E_n$, the wave functions are such that $\psi_{\pm 1}$ vanish asymptotically. Thus, this Wronskian encloses information regarding the quantization condition of this problem.
On the other hand, the remaining object $W_{[0,\pm]}$ depends on the the asymptotic behaviour on the real line as well as near the origin, being associated to the radial problem we are interested in.

Introducing
\vspace{-0.2cm}
\begin{eqnarray}\label{defspecwronsk}
~~~~C(E,\alpha) &=& W_{[-1,1]}(M,\ell,E,\alpha),\\
\nonumber
D^{(\mp)}(E,\alpha) &=& W_{[0,\pm]}(M,\ell,E,\alpha),
\end{eqnarray}

\ni the objects $C(E,\alpha), D(E,\alpha)$ can be written as Wronskians of linearly independent solutions which are entire functions. Thus, these newly defined functions are also entire and we can use the Weierstrass factorization theorem to express the entire spectral determinants as an infinite product.
A zero of $D(E,\alpha)$ indicates that the eigenproblem has a solution for which the eigenfunction corresponding to the eigenvalue $E$ satisfies the correct boundary conditions. Therefore, we can conclude that this coincides with the spectral determinant, up to a normalizing factor, vanishing at the possible eigenvalues of the differential equation.
What is more, the definitions above furnish us with the analogue of Baxter $T$-$Q$ relations (also in \cite{Watkins} and reference [1] therein)
\begin{eqnarray}\label{cdrelation}
\fl\hspace*{-0.5cm}
C(E,\alpha) D^{(\pm)}(E,\alpha) &=&
\omega^{+\frac{2\ell+1+\alpha}{2}} D^{(\pm)}\left(\omega^{+2}E, -\alpha \right) +
\omega^{-\frac{2\ell+1+\alpha}{2}} D^{(\pm)}\left(\omega^{-2}E, -\alpha \right),
\end{eqnarray}

\ni whose zeroes $E=E_j^{(\alpha)}$ are characterised by
\begin{eqnarray}\label{BAEspecdet}
\frac{D^{(\pm)}\left(\omega^{-2}E_j^{(\alpha)}, -\alpha \right)}
{ D^{(\pm)}\left(\omega^{+2}E_j^{(\alpha)}, -\alpha \right)} = - \omega^{(\ele+\alpha)}.
\end{eqnarray}

Part of the solutions for this equation correspond to $C(E,\alpha)$ and the remaining ones are associated to zeros of $D(E,\alpha)$.
Because the problem defined on the real line is Hermitian the zeros of $D(E,\alpha)$ should be real. Therefore, any complex eigenvalue should be a zero of $C(E,\alpha)$ but there is no guarantee that the zeros of $C(E,\alpha)$ cannot be real. 

In order to use the Weierstrass factorization, we must take into account that a simple WKB approximation tells us that the
zeros accumulate towards infinity along the positive real axis according to $E_n \sim n^{\frac{2M}{M+1}}$, so if $M>1$ we can establish a convergent product $D^{(-)} \equiv D$ in the form
\begin{eqnarray}\label{specdetprod}
D(E,\alpha)= D(0,\alpha)\prod_{k=0}^\infty \left(1-\frac{E}{E_k^{(\alpha)}} \right),
\end{eqnarray}

\ni which leads immediately to an expression of the form of the Bethe Ansatz Equations,
\begin{eqnarray}\label{BAEprod}
\prod_{k=0}^\infty \left( \frac{E_k^{({\alpha})} - \omega^{-2}E_j^{({-\alpha})}}{E_k^{({\alpha})} - \omega^{2}E_j^{({-\alpha})}} \right) = - \omega^{(2\ell+1 + {\alpha})}.
\end{eqnarray}

In this case, they correspond to two sets of coupled equations, both for the values of $+\alpha$ and for $-\alpha$. The problems defined for $+\alpha$ and $-\alpha$ have different properties, reflecting on the eigenvalues, or Bethe roots. In the differential equation framework, for one case we have a convex potential whereas for the other a double well potential, respectively \cite{Suzuki-Application}. For $M=1$, we cannot guarantee the convergence of the infinite product above and changes are expected to be necessary. The relations (\ref{solE0a0M1}), though, make it simple to obtain the correct equations for  the $T$-$Q$ and Quantum Wronskian relations.
One should note that, rather than the $Q(E,\alpha)$-operator itself, the spectral determinant $D(E,\alpha)$ should be identified with the quantity $A(E,\alpha)$, defined by
\begin{eqnarray}\label{QABax}
Q\left(  \omega ^{\pm 2} E,-\alpha \right)=
\omega ^{\pm \frac{\ele + \alpha}{2}} A\left(\omega ^{\pm 2}E,-\alpha \right).
\end{eqnarray}

To obtain the correct normalisation factor to the Hadamard factorization we observe that the solution of the differential equation at vanishing value of energy can be found exactly (\ref{solE0}) and by normalising (\ref{asymPsi}) and (\ref{basisorigin}) for instance with $\frac{1}{\sqrt{2i}}$ and 
$\left(\frac{M+1}{2}\right)^{\mp \frac{2 \ell +1}{M+1}}\frac{ \Gamma  \left( \mp \frac{2 \ell +1}{M+1}\right)}{\Gamma \left(\frac{M \mp (2\ell+1) +\alpha +1}{2 (M+1)}\right)}$ respectively, gives
\begin{eqnarray}
\fl\quad
D(0,\alpha) 
\sim (2 \ell +1)  \left(\frac{M+1}{2}\right)^{\frac{2 \ell +1}{M+1}}
\frac{ \Gamma  \left(\frac{2 \ell +1}{M+1}\right)}{\Gamma \left(\frac{M+2 \ell +\alpha +2}{2 (M+1)}\right)}.
\end{eqnarray}
and the expansion of the spectral determinant above (\ref{specdetprod}) has the expected leading behaviour
\begin{eqnarray}\label{leadbehD}\label{leadbehDD}
\fl
\log D(E,\alpha) \sim \frac{\pi}{\cos\left(\frac{\pi }{2 M}\right)}  \left(-\frac{E }{\nu }\right)^{\frac{M+1}{2 M}},
\qquad
  \nu \equiv \( 2\sqrt{\pi} \frac{\Gamma(\frac{3M+1}{2M})}{\Gamma(\frac{2M+1}{2M})} \)^\frac{2M}{M+1},
\end{eqnarray}
well known in the literature.
An important result of the spectral determinant expansion above is the appearance of the constants which are coefficient of the higher order terms in this expansion in powers of the spectral parameter. These correspond to the eigenvalues of the commuting set of local conserved operators in the quantum AKNS model.
The reformulation here presented is well known now and allows one to transform our original differential problems into a set of coupled nonlinear equations. 
In the next section we will show how the coefficients in the expansion of the spectral determinant can be identified as the integrals of motion corresponding to the quantum integrable model.


\section{Spectral Determinants and  Integrals of Motion}

It is an important result that one can transform our initial eigenvalue problem into algebraic equations but  the simplification is only relative as the number of equations to solve becomes infinite. 
In any case, this alternative method which is based not on the differential equation directly but rather in the functional relations one can construct from it
allows one to extend the toolbox to solve the original eigenvalue problem as well as giving access to important information in an exact analytic form. 
In the thermodynamic limit, this approach in essence  transforms the differential problem in nonlinear integral equations, usually referred to as NLIEs.
This has a central role in our strategy to compute nonlocal conserved quantities for the QIM from an analysis of the ODE.


The starting point is to consider the $C$-$D$ relation (\ref{cdrelation}),
which is equivalent to the $T$-$Q$ relation
\begin{eqnarray}\label{TQ-relation}
 T(E ,\alpha )Q(E ,\alpha ) &=&
Q\left(\omega^{2} E,-\alpha \right)+
Q\left(\omega^{-2} E ,-\alpha\right),
\end{eqnarray}

\ni if we identify the spectral determinant $D(E,\alpha)$ with the quantity $A(E,\alpha)$ in (\ref{QABax}).
One then defines a new object,
\begin{eqnarray}
a(E,\alpha) = \omega^{\ele+\alpha}\frac{D(\omega^{+2} E(\alpha),-\alpha)}{D(\omega^{-2}E(\alpha),-\alpha)},
\end{eqnarray}

\ni so that for the zeros $E_k(\alpha)$ of $A(E(\alpha),-\alpha)$ the associated Bethe equation may be recast in the form of
\begin{eqnarray}
a(E_k,\alpha) + 1 =0.
\end{eqnarray}

This is actually the same system of two coupled equations, one for $+\alpha$ and one for $-\alpha$.
We make the assumption that
$T(E,\alpha)$ and $A(E,\alpha)$, or equivalently $C(E,\alpha)$ and $D(E,\alpha)$ respectvely, are entire functions so that we can factorize them in terms of their zeros in a convergent product through a Weierstrass decomposition. Form such a factorisation we can observe that $a\left(E,-\alpha \right) a(E ,-\alpha )^*=1$, if the energy levels are real.


One can write the quantisation condition as
\begin{eqnarray}
\fl
\log a(E,\alpha) = \log \left( \omega^{\ele+\alpha} \right) + \sum_{k=0}^\infty 
\log \left(  \frac{1- \omega^2 \frac{E(\alpha)}{E_k(-\alpha)}}{1- \omega^{-2} \frac{E(\alpha)}{E_k(-\alpha)}} \right)
=2 i \pi   \left(m_k+\frac{1}{2}\right),
\end{eqnarray}

\ni so that for each eigenvalues $E_k$ there are infinite possibilities for the integer parameter $m_k$.
When we take the logarithm of the expression above we introduce an arbitrariness in the choice of the branch, or the Riemann sheet, denoted by an integer $m_k$, in such a way that eigenvalue $E_k$ is associated to one of the integers $m_k$.
A certain eigenstate of the $Q$ operator is characterized by a unique set of integers $\{ m_k \}$.
However, not every set of integers will correspond to an eigenstate of $Q$.

We can replace the infinite sum over zeroes by a contour integral. The total contour can be split in one part containing the positive real line and one contour for the remaining complex plane. Using the reparameterization
\begin{eqnarray}
E =e^{\frac{2M}{M+1}\theta },
\end{eqnarray}

\ni the logarithm expansion of the quantity $a(\theta,\alpha)$ can be expanded as an integral which can be written as a sum of infinitely many residues plus a diverging contribution at the large circle  \cite{DestriDeVega-NewTBA, DestriDeVega-Unified},
%
after the inclusion of a term responsible for the leading behaviour following (\ref{leadbehDD}),
\begin{eqnarray}
\fl
\log a(\theta, \alpha)=
\log\left(\omega ^{\ele + \alpha}\right)
-\frac{2 \pi \imath}{\nu^{\frac{M+1}{2 M}} }e^{\theta } + \\
\fl\nn
+\int_{c_1} K_+(\theta -w) \log (a(w, \alpha)+1) \, dw
-\int_{c_2} K_+(\theta -w) \log \left(\frac{1}{a(w, \alpha)}+1\right) \, dw + \\
\fl\nn
+\int_{c_1} K_-(\theta -w) \log (a(w,-\alpha)+1) \, dw
-\int_{c_2} K_-(\theta -w) \log \left(\frac{1}{a(w,-\alpha)}+1\right) \, dw ,
\end{eqnarray}

\ni where the contours $c_1, c_2$ are placed just above and below the real axis and the exact forms of the kernels $K_{\pm} (\theta)$ were given in integral form in \cite{Suzuki-Fun,DDT-SpecEquiv}.
These integrals can be written as infinite sums, so the kernel may be expressed as
\begin{eqnarray}
K_{\pm} (\theta) = \frac{1}{2} (\varphi (\theta )\pm \chi (\theta )),
\end{eqnarray}
with
\begin{eqnarray}
\fl
\varphi (\theta ) &=&
\sum _{n=1}^{\infty } \frac{M}{\pi}e^{-2 \theta  M n} \tan (\pi  M n)-
\sum _{n=1}^{\infty } \frac{1}{\pi }e^{\theta  (1-2 n)} \cot \left(\frac{\pi  (2 n-1)}{2 M}\right), \\
   \fl
\chi (\theta ) &=&
\sum _{n=1}^{\infty } \frac{M}{2 \pi } e^{\theta  M (1-2 n)} \cot \left(\frac{\pi }{2}  M (2 n-1)\right)-
\sum _{n=1}^{\infty } \frac{1}{2 \pi } e^{-2 \theta  n} \tan \left(\frac{\pi  n}{M}\right),
\end{eqnarray}



Therefore, the sums over the residues only give an asymptotic series expansion and from them we can construct the conserved charges. The NLIE for the KdV  $su(2)$ problem, with vanishing $\alpha$, involves a function which coincides with the solition-solition scattering amplitude for the sine-Gordon model. The same NLIE was earlier derived for the Bethe ansatz equation associated with the XXZ model and by taking the appropriate continuous limit leads to the sine-Gordon field theory. Thus, the integrals of motion of the aforementioned models are related to each other. In this case, the scatering matrices are the ones that should appear in the quantum AKNS field theory and the conserved charges are equivalent in both models.

After some identifications we can relate the integrals of motion to the integrals appearing in the expansion. 
In the case of the nonlocal ones, we have



\begin{eqnarray}
\tilde{I}_{2 n} = -\frac{1}{4 \pi }\left(\frac {M+1}{2^{M}}\right)^{2 n}\frac{M}{\cos(\pi  M n)} \times ~~~~ \\
\nn
\left(
\int_{c_1} e^{2 n M i w} \log \left({a(\alpha ,w)}+1\right) \, dw - \int_{c_2} e^{2 n M i w} \log \left(\frac{1}{a(\alpha ,w)}+1\right) \, dw  \right. ~~~~ \\ 
\nn
\left. 
+\int_{c_1} e^{2 n M i w} \log \left({a(-\alpha ,w)}+1\right) \, dw - \int_{c_2} e^{2 n M i w} \log \left(\frac{1}{a(-\alpha ,w)}+1\right) \, dw
\right),
\end{eqnarray}

\begin{eqnarray}
\tilde{I}_{2 n-1}=-\frac{1}{4 \pi }\left(\frac{M+1}{2^{M} }\right)^{2 n-1}M \cos \left(\frac{\pi  M (2n-1)}{2}\right) \times ~~~~ \\
\nn
\left(
\int_{c_1} e^{(2 n-1) M i w} \log \left({a(\alpha ,w)}+1\right) \, dw - \int_{c_2} e^{(2 n-1) M i w} \log \left(\frac{1}{a(\alpha ,w)}+1\right) \, dw  \right. ~~~~ \\ 
\nn
\left. 
-\int_{c_1} e^{(2 n-1) M i w} \log \left({a(-\alpha ,w)}+1\right) \, dw + \int_{c_2} e^{(2 n-1) M i w} \log \left(\frac{1}{a(-\alpha ,w)}+1\right) \, dw
\right).
\end{eqnarray}


\ni and similar expressions for the local ones \cite{DT-WKB, AD-Exc}. We notice that 

In order to have a consistent notation with respect to the literature it is convenient to make the following identifications
$\tilde{H}_{n}=\tilde{I}_{2 n}$ and $\tilde{H}_{\frac{2 n-1}{2}}=\tilde{I}_{2 n-1}$ for when $\alpha\neq0$.
For vanishing $\alpha$, the contributions of the integrals coincide so the last two terms in the series above cease to exist and one recovers the results of \cite{BLZ-1, BLZ-2, BLZ-3, DT-WKB}.
A WKB analysis of (\ref{genharmoscode}) leads to an expansion of the spectral determinant in terms of local conserved charges which can be found in e. g \cite{BLZ-Higher, DT-WKB} when $\alpha=0$. The local charges for the supersymmetric case are investigated in detail in \cite{AD-Exc} but also have both odd and even contributions, the latter vanishing with $\alpha$, as expected.


The expressions above though provide a recursive scheme which is efficient to make numerical comparisons. It is also an important tool from a practical viewpoint, due to the difficulty of compute the objects $I_n$ and -especially- $\tilde{H}_n$, the local and nonlocal conserved charges of the model.
Having in mind the equivalence between Baxter's $Q(E)$ operator and the spectral determinant $D(E)$ we will show in the next section how the nonlocal integrals of motion can be determined in a reasonable way.


\section{Nonlocal Conserved Charges from the ODE}

We have seen that the supersymmetric model here studied has an important feature, namely, it interpolates the properties of the quantum KdV field theory or, equivalently, that of the sine-Gordon model and the XXZ spin lattice, which are all well understood.  Besides being characterized by local integrals of motion of the kind previously presented for the AKNS analogue, there is another set of important commuting operators, the denoted nonlocal integrals of motion. Originally defined to be the coefficients in a Taylor expansion of the transfer matrix operator, e.g.  \cite{BLZ-2, DestriDeVega-YB},
\begin{eqnarray}\label{nonlocexpsu2}
T(\lambda )=2 \cos (2 \pi  p)+\sum _{n=1}^{\infty} G_n \lambda ^{2 n},
\end{eqnarray}

\ni with $\lambda$ being the spectral parameter related to $E$ according to
$E =
\lambda ^2 \rho$, where $\rho =2^{\frac{2 M}{M+1}} (M+1)^{\frac{2 M}{M+1}} \Gamma \left(\frac{M}{M+1}\right)^2$.

An alternative way of expressing the nonlocal conserved charges, in the case of an underlying $su(2)$ symmetry, consists of defining them as $H_n$ in terms of the coefficients of an expansion of the $Q$-operator in powers of the variable $\lambda   \frac{\Gamma \left(1-\beta ^2\right)}{\beta ^2}$.
The dual nonlocal integrals of motion, specified by $\tilde{H}_n\left(\beta ^2\right)=H_n\left(\frac{1}{\beta ^2}\right)$, give rise to the following expansion of the operator $A(E)$, which corresponds to the spectral determinant  $D(E)$ of the corresponding Schr\"{o}dinger problem. This means we can write down an asymptotic expansion, possibly with limited radius of convergence,
\begin{eqnarray}
\log D(E) \sim \sum_{n=1}^{\infty} \left( \frac{2^M}{M+1} \right)^{2n} \tilde{H}_n (-E)^{-n (M+1)},
\end{eqnarray}
for large values of the spectral parameter $E$.

However, we have seen that the addition of the $\alpha$-term brought another set of local integrals of motion, characterized by even indices, to the pre-existing ones, specified by odd indices only. Therefore we would expect the same to happen with the nonlocal charges. This implies having an expansion
\begin{eqnarray}
T(\lambda,\alpha)=T(0,\alpha) + \sum _{n=1}^{\infty} G_\frac{n}{2} \lambda ^{n},
\end{eqnarray}
which reduces to (\ref{nonlocexpsu2}) when  all even powers of $\lambda$ vanish with $\alpha$.
So the first nonlocal conserved charge $G_\frac{1}{2}$ appears in power of $\lambda$, rather than $G_1$ at order $\lambda^2$, and has no equivalent when $\alpha=0$, unlike $G_1$. Moreover, the dual nonlocal charges also come as twice as many,
\begin{eqnarray}\label{logAonlyH}
\log D(E,\alpha) \sim \sum_{n=1}^{\infty} \left( \frac{2^M}{M+1} \right)^{n} \tilde{H}_{\frac{n}{2}} (-E)^{-n \frac{(M+1)}{2}},
\end{eqnarray}
and the first integral motion in the expansion would have no counterpart in the $su(2)$ scenario and $\tilde{H}_1$ would be the first nonvanishing contribution since no $\tilde{H}_{\frac{1}{2}}$ would be present in that case.

The nonlocal integral equations allow us to compute the charges numerically but analytic expressions are difficult to obtain, particularly in the case of the nonlocal integrals of motion. In \cite{BLZ-1,BLZ-2,BLZ-3} the $G_n$ are formally computed in generality for $\alpha=0$ and exactly evaluated for first two contributions in the expansion. 
It would be in principle advantageous if one could determine these important objects by investigating the associated ODE with nonvanishing $\alpha$ by the existing spectral correspondence.
In fact, here we show that these properties of the operators $A(E,\alpha)$ and $Q(E,\alpha)$ can be obtained in a systematic fashion by investigating the same original differential equation (\ref{genharmoscode}).
Although the construction of the local integrals of motion can done directly from the latter, in the case of nonlocal charges some reparameterizations prove to be very convenient. 
For that, we introduce the new variables,
\begin{equation}
\fl\quad
s=
\log\left( \pm  \varepsilon^{-\frac{M}{M+1}} x \right), \qquad\qquad \Sigma(s)=
\frac{1}{\sqrt{x}}\psi(x),
\end{equation}
where the $\varepsilon$ is related to the spectral parameter through
\begin{eqnarray}
\mathcal{F} = \varepsilon^M = (\pm E)^{-\frac{M+1}{2}}.
\end{eqnarray}
This transforms the problem into an equivalent one, but free from singularities at the origin again,
\begin{eqnarray}\label{nonlocode}
\left[ -\frac{d^2}{ds^2}+ \left( \frac{\ele}{2}\right)^2 \mp e^{2s}+ (\pm)^{M+1} \alpha ~ e^{(M+1)s}\mathcal{F} + e^{2(M+1)s}\mathcal{F}^2 \right]\Sigma(s)=0,
\end{eqnarray}
which corresponds to a potential expanded in terms of powers of $\mathcal{F}$. This can be associated to the AKNS problem by taking $M=\frac{1-\beta^2}{\beta^2}, \ell = 2\frac{P}{\beta^2} -\frac{1}{2}, \alpha = 4\frac{\bar{P}}{\beta^2}$.

For simplicity we can work with the choice $\varepsilon = E^{-\frac{M+1}{2M}}$; otherwise, we must replace
$s$ by $s + \imath \frac{\pi}{2}$, closely related to the associated \pt-symmetric problem. 
This structure tells us that for small values of $\mathcal{F}$, i.e., large values of $E$, this equation can be solved perturbatively, order by order.
The unperturbed potential, with $\mathcal{F}=0$,
which also corresponds to $s \ra - \infty$, has two linearly independent solutions given by Bessel functions,
\begin{eqnarray}
\Sigma_\pm(s)= 
\mathcal{N}_\pm ~
\Gamma \left(1 \pm \frac{\ele}{2}\right) J_{\pm \frac{\ele}{2}}\left(e^{s} \right),
\end{eqnarray}

\ni satisfying required boundary conditions of vanishing at $s=\pm\infty$ and $s=\infty$, respectively, as well as the familiar relations, with $\mathcal{N}_\pm =\pi^{\pm \frac{1}{2}} \Gamma \left(\frac{2 \ell +1}{2}+\frac{1\pm 1}{2}\right)^{\mp 1}$,
\begin{eqnarray}\label{specdetnonlocexp}
\nn
W[\Sigma_-(s),\Sigma_+(s)] &=&  2, \\
~~~~~~~~~~~~\Sigma(s) &=& \frac{1}{2}D^+(E,\alpha) \Sigma_+(s) + \frac{1}{2}D^-(E,\alpha) \Sigma_-(s).
\end{eqnarray}
In fact, the solutions $\Sigma_{\pm}(s)$ have first order contribution of the type $\varphi^{\pm}(x)$ in (\ref{basisorigin}), after the appropriate change of variables.

As $s \rightarrow - \infty$, we have that $\Sigma_+(s)\rightarrow 0$, so we can establish the proportionality of the spectral determinant
$D(E,\alpha) = D^{(-)}(E,\alpha) = W_{[0,+]}(E,\alpha)$ with the wavefunction in this limit.
However, expression (\ref{specdetnonlocexp}) is generally true in this regime and does not rely on any interpretation of the $D^{\pm}(E,\alpha)$ as spectral determinants which vanish for appropriate boundary conditions. In fact, in the present context the transformed differential equation does not exactly constitute an eigenvalue problem, albeit closely related to one.
Therefore, these Wronskians are simply the projection of a solution in the aforementioned basis and the problem of determining them relies on a solution of the differential equation. 

Because the latter can be seen as a perturbed equation in powers of the parameter $\mathcal{F}$ we can assume a general solution of the form
\begin{eqnarray}\label{sigmaexpup}
\Sigma(s) = \Sigma^{(+)}(s)+\Sigma^{(-)}(s), \\
\Sigma^{(\pm)}(s)=\Delta _0^{(\pm)}(s)+\sum _{j=1}^{\infty } \mathcal{F}^j \Delta _j^{(\pm)}(s), 
\end{eqnarray}
where in lowest order $\Delta _0^{(\pm)}(s) =  \Sigma_{\pm}(s)$ is the solution of the unperturbed equation ($\mathcal{F}=0$) satisfying the necessary boundary conditions.
Besides, we have added all the remaining perturbative contributions so that the result is expressible as a series in integer powers of $\mathcal{F}$.
Restricting for now our analysis to the ground state of the $Q$-operator for the perturbed Hamiltonian (\ref{nonlocode}) we can expand the wavefunction in powers of $\mathcal{F}$ and solve the associated Sch\"{o}dinger equation perturbatively at each order, first,
\begin{eqnarray}
\fl
\Delta _1^{(\pm)}(s)=
&\mp&
\alpha\frac{ \Gamma \left(1\pm \frac{2 \ell +1}{2}\right)^2}{2 \ell +1}
\Sigma_\mp(s)
\int_{s_{11}}^s e^{(M+1) y} J_{\pm\frac{2 \ell +1}{2}}\left(e^y\right){}^2 \, dy
+ \\
\fl
\nonumber
 &\pm&
   \frac{\alpha}{2} \frac{\pi}{ \cos (\pi  \ell )}
   \Sigma_\pm(s)
   \int_{s_{12}}^s e^{(M+1) y} J_{-\frac{2 \ell +1}{2}}\left(e^y\right) J_{\frac{2 \ell +1}{2}}\left(e^y\right) \, dy
\end{eqnarray}

\ni then, by solving the resulting equation as the coefficient of $\mathcal{F}^2$,
\begin{eqnarray}
\fl
\Delta _2^{(\pm)}(s)=& \mp &
\frac{\Gamma \left(1\pm \frac{2 \ell +1}{2} \right)^2 }{2 \ell +1} 
\Sigma_\mp(s)
\int_{s_{21}}^s e^{2 (M+1) y} J_{\pm\frac{2 \ell +1}{2}}\left(e^y\right){}^2 \, dy+
\\
\fl
\nonumber
&-&
\frac{\alpha}{2}  \Gamma \left(\pm\frac{2 \ell +1}{2}\right) 
\Sigma_\mp(s)
\int_{s_{21}}^s e^{(M+1) y} \Delta _1(y) J_{\pm\frac{2 \ell +1}{2}}\left(e^y\right) \, dy
+ \\
\fl
\nonumber
&\pm&
   \frac{1}{2} \frac{\pi}{\cos (\pi  \ell )}
   \Sigma_\pm(s)
   \int_{s_{22}}^s e^{2 (M+1) y} J_{-\frac{2 \ell +1}{2}}\left(e^y\right) J_{\frac{2 \ell +1}{2}}\left(e^y\right) \, dy +  
   \\
   \fl
   \nonumber
   &-&
   \frac{\alpha }{2} \Gamma \left(\mp\frac{2 \ell +1}{2}\right)    
   \Sigma_\pm(s)
   \int_{s_{22}}^s e^{(M+1) y} \Delta _1(y) J_{\mp\frac{2 \ell +1}{2}}\left(e^y\right) \, dy
\end{eqnarray}

\ni and so on, successively, for higher orders. We see that $\Delta_2(s)$ depends explicitly on $\Delta_1(s)$ and the same pattern occurs at all orders: to determine the solution at each order it is necessary to know the solutions at previous ones.
In fact the solution for a generic order can be found in closed formula described by the expression,
\begin{eqnarray}\label{DeltaK}
\fl
\nonumber
\Delta _{k}^{(\pm)}(s)&=& - 
\frac{\Gamma \left(\frac{2 \ell +1}{2} \right)}{2} 
\Sigma_-(s)
\int_{s_{k1}}^s e^{(M+1) y} J_{\frac{2 \ell +1}{2}}\left(e^y\right) 
\left(  e^{(M+1) y} \Delta_{k-2}^{(\pm)}(y)+ \alpha  \Delta_{k-1}^{(\pm)}(y) \right) \, dy 
\\
\fl
\nonumber
&-&
   \frac{\Gamma \left(-\frac{2 \ell +1}{2} \right)}{2}
   \Sigma_+(s)
   \int_{s_{k2}}^s e^{(M+1) y} J_{-\frac{2 \ell +1}{2}}\left(e^y\right) 
   \left( e^{(M+1) y} \Delta_{k-2}^{(\pm)}(y)+ \alpha  \Delta_{k-1}^{(\pm)}(y)  \right) \, dy   .
\end{eqnarray}
The constants of integration are appropriately fixed to satisfy the required boundary conditions so that contributions which do not behave appropriately are excluded form $\Delta_k(s)$.

When $\alpha$ vanishes, the contribution of $\Delta _1(s)$ is reabsorbed in the unperturbed solution.
Curiously, the contribution $\Delta _2(s;\alpha=0)$ is almost the same as $\Delta _1(s;\alpha=1)$ and that is because
$V \sim \alpha~ e^{(M+1) s} \mathcal{F} +  e^{2 (M+1) s} \mathcal{F}^2 $, i.e. when the supersymmetric parameter vanishes the first contribution, coming at the order 
$\mathcal{F}^2$, is the square of the first contribution when such parameter is nonzero, at first order in $\mathcal{F}$.

The expressions above seem cumbersome but from them we can see that it is convenient to introduce the following two quantities,
\begin{eqnarray}\label{integralJ}\label{integralK}
\fl\nn
~~~~\mathcal{J}_n (y)
&=&   \frac{ _2F_3\left(\ell +1,\frac{n}{2}(M+1)+\elef;\elef+1,\frac{n}{2}(M+1)+\elef +1,2 (\ell+1);-e^{2y}\right)}
   {2^{2 \ell +1} e^{-2((M+1) n+2 \ell+1)y}((M+1) n+2 \ell +1)\Gamma \left(\elef+1\right)^2 },\\
\fl
~~\mathcal{K}_n (y)
&=&   \frac{ _2F_3\left(\frac{1}{2},\frac{n}{2}(M+1);\frac{n}{2}(M+1)+1,1-\elef ,1+\elef;-e^{2y}\right)}{(M+1) n
   \Gamma \left(1-\elef \right) \Gamma \left( 1+ \elef \right)} \, e^{2(M+1) n y} ,
\end{eqnarray}
which arise as the indefinite integral forms involving $J_{\frac{2 \ell +1}{2}}\left(e^y\right)^2$ and $J_{\frac{2 \ell +1}{2}}\left(e^y\right)J_{-\frac{2 \ell +1}{2}}\left(e^y\right)$ in the expressions for $\Delta_n(s)$ above. The equivalent result for $J_{-\frac{2 \ell +1}{2}}\left(e^y\right)^2$ can be trivially obtained by just replacing $\ell \ra -(\ell+1)$ in the former.
Taking the lower integration limit to be $x_i = \infty$, or $s_i = \infty$, and evaluating it near the origin, $x=0$ or $s = - \infty$, as this is the region we are interested in the wavefunction, we obtain the quantities
\begin{eqnarray}
\fl
J_n &=&
   \frac{\Gamma \left(\frac{1}{2}-\frac{n}{2} (M+1)\right) \Gamma \left(\elef+\frac{n}{2} (M+1)\right)}{2 \sqrt{\pi } \Gamma
   \left(1-\frac{n}{2} (M+1)\right) \Gamma \left(\elef-\frac{n}{2} (M+1)+1\right)},\\
\fl
K_n &=&
\frac{\Gamma \left(\frac{n}{2} (M+1)\right) \Gamma \left(\frac{1}{2}-\frac{n}{2} (M+1)\right)}
{2 \sqrt{\pi } \Gamma \left(1-\elef-\frac{n}{2} (M+1)\right) \Gamma \left(1+\elef-\frac{n}{2} (M+1)\right)}.
\end{eqnarray}
These are related to each other in a very simple way,
\begin{eqnarray}
\nonumber
\frac{J_n}{K_n}= \frac{ \cos 
\frac{\pi}{2}   \left((M+1) n-1\right)~~~~~~~~~~~~~~
}{ \cos 
\frac{\pi}{2}   \left((M+1) n-1+(2 \ell+1)\right) 
},
\end{eqnarray}
and can be easily related to the quantity $\tilde{I}_2=\tilde{H}_1$ calculated in \cite{BLZ-2},
\begin{eqnarray}
J_2 = \frac{2}{\pi} \sin{M\pi} \left( \frac{2^M}{M+1} \right)^2 \tilde{H}_1,
\end{eqnarray}
with the $\sin$ contribution arriving because we are expanding in $\mathcal{F} = E^{-\frac{M+1}{2}}$.
These constants play an important role in the theory as they will then appear when one is evaluating the solutions at each order. 

We have seen that in every order there are contributions in the direction of both $\Sigma_\pm(s)$ and one can collect them and determine an expansion for the then called spectral determinant.
This implies that, up to the Weierstrass normalisation factor which fixes $D(0,\alpha)$, an expansion of the spectral determinant in the form
\begin{eqnarray}\label{specdetnonlocexpi}
D(E,\alpha) \sim 1+ \sum_{n=1}^\infty F_n \, \mathcal{F}^n, \qquad
\textrm{where} \qquad
F_n = \lim_{s \ra - \infty} \frac{\Delta_n^{(+)}(s)}{\Delta_0^{(-)}(s)}.
\end{eqnarray}

In this limit we are interested in, the above expressions allow us to compute the quantity $F_1$, appearing as the coefficient in the first order expansion in terms of $E^{-\frac{M+1}{2}}$ and it turns out to be given by a particularly simple expression which is proportional to the constant $J_1$ we have just defined,
\begin{eqnarray}\label{NLF1}
F_1 =  \alpha \frac{\pi}{2} J_1.
\end{eqnarray}
This object has a simple form and does not contribute in the $su(2)$ limit because it is proportional to the supersymmetric parameter $\alpha$.

Solving the second order term is trickier as it depends on the solution $\Delta _1(s)$, which is to be integrated in the object
\begin{eqnarray}
B_1(s) = \int_{s_0}^s e^{(M+1) y} \Delta _1(y) J_{\pm \frac{2 \ell +1}{2}}\left(e^y\right) \, dy.
\end{eqnarray}
The evaluation in this case,  when $\lim_{s \ra -\infty}$, show that the factor $F_2$, although having a more complicated form, 
once more the solution can be written down as a function of the fundamental  quantities $J_n$, or $K_n$.
It yields that 
\begin{eqnarray}\label{NLF2}
F_2 &=&  \frac{\pi }{2} J_2 + 
\alpha 
\frac{
\Gamma \left(\frac{1}{2}-\ell \right)
\Gamma \left(\ell +\frac{1}{2}\right)
\Gamma\left(\ell +\frac{3}{2}\right)^2
}{4 \pi ^2 (2 \ell +1)} 
\left(b_1+b_2\right),
\end{eqnarray}
where the remaining constants $b_i$, which arise from the evaluation of $B_1(s)$ at $-\infty$, involve functions of the kind $_5F_4$,
\begin{eqnarray}
b_1 &=& \frac{2 \Gamma (-M) \Gamma (M+\ell +1)}
{M \Gamma \left(\frac{1}{2}-M\right) \Gamma (-M+\ell +1)} \, _5F_{4,1}, \\
\fl
\nn
b_2 &=& \frac{\pi
   \Gamma (M+1) \Gamma (-M-\ell -1)}
   {4^{\ell } (M+2 \ell +2)^2 \Gamma \left(\ell+\frac{3}{2}\right)^2 \Gamma (-M-2 \ell -2) \Gamma \left(-M-\ell -\frac{1}{2}\right)} \, _5F_{4,2},
\end{eqnarray}
and we have used for short the quantities $_5F_{4,1}$ and $_5F_{4,2}$, respectively as
\begin{eqnarray}
\nn\hspace*{-0.5cm}
&& _5F_4\left(\frac{1}{2},-M,-\frac{M}{2},-\ell ,\ell +1;\frac{1}{2}-M,1-\frac{M}{2},-M-\ell ,-M+\ell+1;1\right), \\
\nn\hspace*{-0.5cm}
&& _5F_4\left(M+1,\ell+1,\frac{M}{2}+\ell +1,M+\ell +\frac{3}{2},M+2 \ell +2;\ell +\frac{3}{2},\frac{M}{2}+\ell +2,M+\ell +2,2 \ell +2;1\right).
\end{eqnarray}
The appearance of these hypergeometric functions already in $\tilde{H}_1$ make evident the complications presented by the introduction of the supersymmetric term since for $\alpha=0$ such contributions would appear at higher orders, starting at $\tilde{H}_2$, or $G_2$ as in \cite{Watkins}.
The present results, though, do not rely on any knowledge concerning the properties of Zeta functions.



Now we are finally in a position to evaluate the first few terms in the expansion coming from (\ref{specdetnonlocexpi}),
so that the logarithm expansion of the spectral determinant takes the form
\begin{eqnarray}\label{logDNL}
\log D(E,\alpha) \sim \mathcal{F} {F_1} &+& \mathcal{F}^2 \left( {F_2}- \frac{F_1^2}{2} \right)+\\
\nn
+\mathcal{F}^3\left(F_3 + \frac{F_1^3}{3}-F_2 F_1\right)
&+&\mathcal{F}^4\left(F_4-\frac{F_2^2}{2}-F_3F_1+F_2 F_1^2-\frac{F_1^4}{4} \right) +
\ldots
\end{eqnarray}
In the limit of vanishing $\alpha$, the odd (or half integer, according to other notations) components vanish, as expected from the $su(2)$-symmetric theory, for which only even (or integers, alternatively) terms are present.
For this situation, the first contribution is given simply by
\begin{eqnarray}
F_2 = (-1)^{-(M+1)}\left( \frac{2^M}{M+1} \right)^2 \tilde{I}_2,
\end{eqnarray}

\ni with the standard identification
\begin{eqnarray}
\tilde{I}_2 = \tilde{H}_1=
(M+1)^2\Gamma (-(2 M+1))
\frac{\Gamma (M+1)}{\Gamma (-M)}
\frac{\Gamma \left(\frac{2 \ell +1}{2} +M+1\right)}{\Gamma \left(\frac{2 \ell +1}{2}-M\right)},
\end{eqnarray}

\ni in agreement with (\ref{logAonlyH}). In possession of the expressions (\ref{specdetnonlocexpi}) we can keep on with (\ref{DeltaK}) and determine higher contributions to the solution expansion and therefore extract higher nonlocal integrals of motion. Those terms however become cumbersome with simplifications when $\alpha=0$, as even $F_n$ disappear form (\ref{logDNL}).

Nevertheless, it would be interesting to verify if the method currently under investigation provides any information about the integrals of motion evaluated away from the vacuum of the $Q$ operator. It was shown in \cite{BLZ-Higher} that higher level eigenvalues of the Baxter operator can be obtained by adding to the potential poles located at $x^{2(M+1)}= z_k$. For the first excited state, for instance, 
\begin{equation}\label{pole1}
z_1 = \frac{(2 \ell +1)^2-4 M^2}{M}.
\end{equation}
The corresponding part that needs to be added to (\ref{nonlocode}) is also expressed as a power series in the parameter $\mathcal{F}$,
\begin{eqnarray}
U_L(s) =  4(M+1)\sum_{k=1}^L \sum_{j=1}^\infty \frac{(2 j (M+1)-1)}{z_k^{j}} e^{2 s (j M+j-1)} \mathcal{F}^{\frac{2 (j M+j-1)}{M+1}},
\end{eqnarray}
but the expansion is in rational, not integer, powers - unless $M=1$. Therefore one needs to accommodate this feature in the solution series according to
\begin{eqnarray}\label{sigmaexpupsilon}
\Sigma (s)=\Delta _0(s)+\sum _{j=1}^{\infty } \mathcal{F}^j \Delta _j(s)+\sum _{j=1}^{\infty } \mathcal{F}^{\frac{2 (j M+j-1)}{M+1}} \Upsilon_j(s),
\end{eqnarray}

\ni where $\Upsilon _j$ are the contributions due to the presence of the higher level potential ($L\neq0$), involving non-integer powers of the perturbation parameter.
Here we simplify the notation by eliminating the upper indices $^{(\pm)}$ and present only the positive ones, reminding the negative ones are related to these by a simple change in the sign of the combination $\frac{\ele}{2}$. 

In principle, there are infinitely many $\Upsilon_j$ because this is the expansion of the $U_L$ potential and investigating the behaviour of the this introduction, in a similar way as we have just done for $L=0$, consists of finding solutions to the $\Upsilon_j(s)$ in the expansion (\ref{sigmaexpupsilon}).
It turns out that these terms can be solved in closed form for a generic $j$-th coefficient,
\begin{eqnarray}
\fl
\nonumber
\Upsilon_j(s) &=& 
\Sigma_-(s)
{(2 j (M+1)-1)}(M+1)(2 \ell +1)\Gamma \left(\frac{2 \ell +1}{2} \right)^2 \times 
\\
\fl\nn
&\times&  
\sum _{k=1}^{L} \frac{1}{z_k^{j}} \; \int_{s_{j1}}^s e^{2 y (j (M+1)-1)} J_{\frac{2 \ell +1}{2} }\left(e^y\right){}^2 dy  
\\
\fl
\nonumber
   &+& 
\Sigma_+(s)
{(2 j (M+1)-1)}(M+1)\frac{2\pi}{\cos \left( \pi \ell \right)} \times 
\\
\fl
&\times& 
 \sum _{k=1}^{L} \frac{1}{z_k^{j}} \;  \int_{s_{j2}}^s e^{2 y (j (M+1)-1)} J_{\frac{2 \ell +1}{2} }\left(e^y\right)J_{-\frac{2 \ell +1}{2} }\left(e^y\right)dy 
.
\end{eqnarray}

Once more, in order to determine the contribution of the excited states of level $L$ to spectral determinant expansion we must evaluate the wavefunction in limit where we take $s \ra -\infty$. In this case, the integrals, which have a similar form as those in (\ref{integralJ}) and (\ref{integralJ}), can be evaluated. With this expression at hand we have determined how the introduction of the poles $z_k$ in the ordinary differential equation affects the spectral determinant expansion, 
\begin{eqnarray}
\fl
\nonumber
\Upsilon_j(s) &=& 
\Gamma \left(1-\frac{2 \ell +1}{2} \right) J_{-\frac{2 \ell +1}{2}}\left(e^s\right)
\frac{(2 j (M+1)-1)}{z_k^{j}}(M+1)(2 \ell +1)\Gamma \left(\frac{2 \ell +1}{2} \right)^2 \times \\
\fl\nn
&\times&
\frac{1}{2 \sqrt{\pi }}
\frac{\Gamma \left(\frac{1}{2} (3-2 j (M+1))\right)}{\Gamma \left(\frac{1}{2} (3-2 j(M+1))+\frac{1}{2}\right)}
\frac{\Gamma \left(\frac{1}{2} (2 j (M+1)-3)+\ell +1\right)}{\Gamma \left(\frac{1}{2} (3-2 j (M+1))+\ell +1\right)}
\sum _{k=1}^{L} \frac{1}{z_k^{j}}  \\
\fl
\nonumber
   &+& 
\Gamma \left(1+\frac{2 \ell +1}{2} \right) J_{\frac{2 \ell +1}{2}}\left(e^s\right)
\frac{(2 j (M+1)-1)}{z_k^{j}}(M+1)\frac{2\pi}{\cos \left( \pi \ell \right)} \times \\
\fl
&\times&
\frac{1}{2 \sqrt{\pi }}
\frac{\Gamma \left(\frac{1}{2} (3-2 j (M+1))\right)}{\Gamma \left(\frac{1}{2} (3-2 j (M+1))-\ell \right)}
\frac{\Gamma \left(\frac{1}{2} (2 j (M+1)-2)\right)}{\Gamma \left(\frac{1}{2} (3-2 j (M+1))+\ell +1\right)}
\sum _{k=1}^{L} \frac{1}{z_k^{j}} ,
\end{eqnarray}
evaluated at $s=-\infty$.
Therefore, the expansion suggested allows us to know the exact form of each of the higher level coefficients in the expansion.
The complete evaluation of the function above requires, though, knowledge about the exact positions of the poles, which is known exactly for $L=1$.
However, even if we include the  expression for $z_1$ the overall contribution of the $\Upsilon_j$ does not appear in $D(E,\alpha)$ in the precise form one is interested, namely integer powers of $\mathcal{F}$.
This prevents us from determining the effect of excited states on the nonlocal integrals of motion, indicating the above parameterization is not good to investigate the effects of higher levels $L \geq 0$.

The situation is different when $M=1$, with the introduction of poles contribute to lower orders according to
\begin{eqnarray}
U_L(s) =   \sum_{k=1}^L \frac{24}{z_k} \, e^{2s} \, \mathcal{F}+ 
\sum_{k=1}^L \frac{56}{z_k^2} \, e^{6s} \, \mathcal{F}^3+ \cdots ,
\end{eqnarray}
so that the lowest contribution comes simply as a shift in the term of order $\mathcal{F}$ in (\ref{nonlocode}).
In this situation, it is enough to make the following replacement in $F_1$ and $F_2$, (\ref{NLF1}) and  (\ref{NLF2}), respectively, 
\begin{eqnarray}
\alpha \longrightarrow 
\sum_{k=1}^L \frac{24}{z_k},
\end{eqnarray}
which for the first excited state can be determined analytically with (\ref{pole1}).
If the location of the poles $z_k(\alpha)$ are known for when $\alpha \neq 0$, then in order to evaluate the first contribution of the excited states one needs to take $\alpha \longrightarrow \alpha + \sum_{k=1}^L \frac{24}{z_k(\alpha)}$ instead.

The fact that the poles are located away from the origin guarantees a well behaved modification to the supersymmetric parameter.
Higher contributions can also be calculated but to the best of our knowledge these objects have not been explicitly presented for a direct comparison.
In fact the study of excited states deserves separate attention.
The appearance of the new terms to the original potential at orders $\mathcal{F}^3, \mathcal{F}^5, \cdots$ \, does not allow us to extract the higher level contributions from the vacuum in a simple way. The solution for these cases is obtained by identifying  $\Upsilon_j(s)$ with $\Delta_j(s)$ and following the same procedure for the modified equation at each order.

\section{Conclusions and Perspectives}

The spectral equivalence between $su(2|1)$ supersymmetric spectral determinants {$D(\lambda)$} associated to a class of ordinary differential equations and the eigenvalues of the Baxter {$\hat Q(\lambda)$}-operator of the quantum AKNS field theory has been investigated. 
We have presented a framework capable of proving us with the nonlocal conserved charges associated to the quantum model, $\tilde{H}_n$ in order to determine the whole NLIE expansion together with the local conserved quantities $I_n$.
This constitutes an alternative and efficient method which allows one to determine both odd and even charges, which are missed out from the spectral Zeta function approach of \cite{Watkins}, and extend the results in \cite{DT-WKB, FL-AKNS} for a supersymmetric potential in an elegant way which can be reduced to the previous case when the deformation parameter, responsible for taking an $su(2)$-symmetric problem into an $su(2|1)$ analogue, is set to vanish. 

Many questions remain open, especially regarding the understanding of the higher level IM/ODE correspondence for algebras beyond the local integrals of motion associated to $su(2)$ and $su(2|1)$ \cite{BLZ-Higher, AD-Exc}.
Here a first step is taken in order to also accommodate the nonlocal conserved quantities.
The introduction of poles in the complex plane may destroy in some cases the reality of the spectrum for a certain ordinary differential equation. Therefore, the excitation of higher {$\hat Q(\lambda)$}-states may be seen as a mechanism of breaking of $\mathcal{PT}$-symmetry for a non-Hermitian Hamiltonian, a phenomenon which deserves attention on its own.
Once we have established this equivalence between certain two systems also for the excited states, information extracted in one side may be valuable on the other.


\vspace{1cm}

\section*{Acknowledgments}
I would like to thank Clare Dunning, Joe Watkins and Giota Adamopoulou for useful discussions as well as EPSRC (EP/G039526/1) and SISSA/INFN-Trieste for financial support.

\bibliographystyle{unsrt}

\bibliography{reference}

\end{document}